\def\etal{{\it et al.}\ }
\def\eg{{\it e.g.,}\ }
\def\ie{{\it i.e.,}\ }

\def\kms{\ifmmode{\hbox{km~s}^{-1}}\else{km~s$^{-1}$}\fi}
\def\la{\mathrel{\hbox{\rlap{\hbox{\lower4pt\hbox{$\sim$}}}\raise1pt\hbox{$<$}}}}
\def\ga{\mathrel{\hbox{\rlap{\hbox{\lower4pt\hbox{$\sim$}}}\raise1pt\hbox{$>$}}}}

\documentstyle[aaspp4]{article} 
\begin{document}
\title {\bf Evidence for Stellar Streaming in the Cores of Elliptical Galaxies: \\
A Kinematic Signature of Mergers?}

\author{Michael J. Pierce and Robert C. Berrington}

\affil {Department of Astronomy, Indiana University,
Swain West 319, Bloomington, IN 47405 \\
e-mail: mpierce@astro.indiana.edu, rberring@astro.indiana.edu}

\begin{abstract}

We present evidence for non-Gaussian velocity fields within the 
cores of luminous elliptical galaxies.  This evidence is based 
upon high signal-to-noise, medium-resolution spectroscopy of the 
cores of early-type members of the Virgo and Coma clusters 
obtained with the WIYN 3.5-m telescope.  The Virgo data were 
acquired using an integral-field unit (DensePak) allowing the 
velocity field to be sampled over a variety of spatial scales.  
The Coma data were obtained through single, 2-arcsec diameter 
fibers.  The cross-correlation profiles of luminous ellipticals 
show considerable structure, often having several features with 
amplitudes as high as 10\% that of the cross-correlation peak 
itself.  This structure is most obvious within a radius of $1.5$ 
arcsec (at Virgo), or $\leq$ 100 pc, and is nearly undetectable 
when the data are binned over R $< 15$ arcsec, or $\leq$ 1 kpc.  
Similar features are found in the single-fiber spectra of the 
luminous ellipticals in the Coma Cluster suggesting they are 
ubiquitous to giant ellipticals.  Interesting, only the most 
luminous elliptical galaxies show this phenomena; the central 
regions of lower luminosity ellipticals have regular, Gaussian-like 
profiles.  We interpret this kinematic structure as ``stellar 
streaming'' and suggest that this phenomena could be a relic 
signature of the merger history of luminous elliptical galaxies.  

\end{abstract}


\section{Photometric Properties of Elliptical Galaxies}

It is generally accepted that elliptical galaxies are the merger product 
of smaller mass systems (Toomre 1977).  While elliptical galaxies populate a 
``fundamental plane'' in their global scaling properties 
(\eg Dressler \etal 1987; Djorgovski \& Davis 1987), they also appear 
to be broadly composed of two families, giants and dwarfs 
(\eg Kormendy \& Djorgovski 1989).  These distinct properties are often 
assumed to be a result of luminous ellipticals forming from lower 
luminosity systems through a merger process.  Recently, Faber \etal (1997) 
have shown that the photometric profiles of the cores of elliptical galaxies 
can also be broadly classified into two groups, those with ``cores'' and those 
with ``cusps''.  Systems with cusps have surface brightness profiles which 
continue to rise into their inner-most regions.  In contrast, systems with 
cores have an inner region of almost constant surface brightness with a 
well-defined radius where the profile breaks to form the standard $R^{1/4}$ 
profile.  Core profiles are found exclusively within giant ellipticals 
while those systems with photometric cusps are invariably lower luminosity 
systems.  Faber \etal (1997) suggested that the distended cores of the 
higher luminosity systems could be the result of a kinematic heating of 
the stellar population supplied by massive BHs which survive the 
merger process.  In this scenario, these massive black holes may 
produce detectable stellar ``wakes'' as they traverse the background 
sea of relatively low mass stars.  

We have found evidence for non-Gaussian velocity fields within the 
cores of giant ellipticals. In \S 2 we describe our observational 
data and analysis procedures.  In \S 3 we introduce a ``structure 
index'' in order to quantify the kinematic structure we find.  We 
present evidence that the structure index is strongly 
correlated with the absolute magnitude of the galaxy.  We discuss 
some possible interpretations of this phenomena in \S 4 and \S 5 
contains a summary of our results.

\section{Spectroscopy of Elliptical Galaxies}

High signal-to-noise, medium resolution spectroscopy of several 
hundred elliptical galaxies has been obtained with the WIYN 3.5-m 
telescope on Kitt Peak as part of an extensive investigation of 
the fundamental plane of early-type galaxies.  The sample 
presented in this paper includes the brighter elliptical galaxies in the 
Virgo and Coma clusters.  The Virgo data were acquired with 
``DensePak'' in order to sample these nearby systems over similar 
spatial scales as do 2-3 arcsec fibers with more distant galaxies. 
DensePak is an integral-field unit consisting of a $7 \times 13$ 
array of 3 arcsec diameter fibers which sample a $30 \times 45$ 
arcsec region on the sky (Barden, Sawyer \& Honeycutt 1998) \markcite{bar}.  
A second fiber bundle consisting of 96, robotically positioned, 
2-arcsec diameter fibers (Hydra) was used to acquire spectroscopy 
of the Coma galaxies.  

The exposure times using DensePak were 
typically 30-min for each Virgo galaxy while the Coma data constituted 
4 1-hour exposures with Hydra.  
The signal-to-noise was typically 
about 50 per pixel.  
Both fiber bundles feed a versatile, bench-mounted spectrograph 
enabling a wide variety of configurations.  We chose the standard 
f/6 paraboloid collimator with the refractive f/1.43 ``red'' 
camera and the 860 l/mm grating in second order to produce a 
reciprocal dispersion of 19.9 ~\AA/mm or 0.477 ~\AA/pixel with 
the Tektronics 2048 $\times$ 2048, thinned CCD (T2KC).  This 
configuration produced a FWHM of 2.5 pixels as measured from the 
comparison lines corresponding to an instrumental velocity 
resolution (1 $\sigma$) of 29 \kms.  We chose the spectral region 
from $\lambda\lambda$ 4813\AA~ $-$ 5793\AA~ including H$\beta$, 
Mg-b, and Fe + CaI.  Comparison spectra of a CuAr lamp were taken 
periodically over the course of the night.  Flat field calibration 
was done on each night using a quartz lamp in order to correct 
for the fiber-to-fiber sensitivity variations and to define the 
extraction regions for the individual spectra.  
Observations of several ``super metal-rich'' K-giant stars 
(Faber \etal 1985) were obtained through different fibers over the 
course of each night.  In the case of the DensePak data, we acquired 
these template stars both in and out of focus in order to test for 
artifacts introduced by fiber-to-fiber 
variations in the spectrograph.  We also acquired exposures of 
the twilight sky on each night.  The only differences discernible 
between any of the template and/or twilight exposures were very 
small and negligible focus variations along the spectrograph slit.

All the spectra were processed and extracted using the ``DoHydra'' script 
within IRAF.  The flat-field exposures were used to define the apertures 
for optimal extraction.  CuAr comparison spectra were extracted 
and fitted to provide the wavelength calibration.  Sky spectra were 
extracted and co-added.  For the Hydra data we allocated 25 fibers to 
the sky but in the case of DensePak only 4 fibers are available for sky.  
The calibrated and extracted spectra were saved in a ``multispec'' 
format for later analysis.  We began by computing the median intensities 
of each of the DensePak spectra.  
From these we calculated the intensity-weighted, spatial centroid of the 
galaxy on the DensePak array.  Given the projected position of each fiber 
on the sky, the spectra were binned over various spatial scales 
to create summed object spectra.  The spectra of the template K giant 
stars were extracted and cross-correlated with each object 
spectra using the FXCor task within IRAF.  This task is based upon 
the algorithm of Tonry \& Davis (1979).  The cross-correlation is 
performed in Fourier space but any model fitting is done in real space. 
One advantage of this approach is that the resulting profile is easily 
compared to a model, in our case a Gaussian, and any deviations are readily 
apparent.

\section{Results}

Upon examination of the cross-correlation profiles of the Coma data 
we were immediately struck by the presence of strong irregularities 
in the profiles of the most luminous systems.  The DensePak data for 
Virgo were found to show similar features.  We show some selected 
examples in Figure 1 along with the best-fit Gaussian models.  The 
Coma data were taken over the course of several nights and the data 
from each night was reduced independently.  We found that the features 
present in the cross-correlation peak repeated from night-to-night.   
We tried several different K-giant stars as templates but could 
not find any detectable differences in the cross-correlation profiles.  
For the DensePak data nearly all 
the Virgo galaxies were centered on the same fiber of the array.  
We did find that when the DensePak data were binned over successfully 
larger radii the amplitude of the deviations decreased rapidly.  
Over scales of $R < 15$ arcsec, or about 1 kpc, the cross-correlation 
profiles became highly Gaussian.  While it is possible that small 
asymmetries in the profiles could be introduced by line-strength variations 
it seems implausible that such variations can produce the dramatic 
substructure seen in Figure 1.  We therefore conclude that the features 
we find in the cross-correlation profiles are real and due to distinct 
kinematic components in the velocity field of these galaxies.  

In an attempt to quantify the degree of substructure in the cross-correlation 
profiles we defined a ``structure index'' (S).  Using the Coma data 
taken over different nights as a guide we adopted criteria as to what 
constituted a significant deviation from the best-fit Gaussian profile.  
Deviations below the full width at half maximum (FWHM) of the profile 
were ignored as were any deviations which were centered on the profile 
peak.  We then simply counted the deviations, both negative and positive, 
from the best fitting Gaussians.  The value of S was estimated by both of 
us and any discrepancies were resolved.  Figure 2 shows the value of S 
for each galaxy plotted against its absolute magnitude in B.  The apparent 
B magnitudes were taken from the RC3 (de Vaucouleurs \etal 1991) or 
computed from our unpublished I-band photometry assuming a mean $B-I$ 
color of 2.35.  For the Virgo sample we assumed an average distance 
modulus of 31.0 and for the Coma sample we assumed an average distance 
modulus of 34.9.  The spectroscopic and photometric data will be presented 
in greater detail at a later date.  Figure 2 shows a clear trend of S 
increasing with the absolute magnitude of the galaxy for those systems 
with $M_B \leq -20.0$.  Lower luminosity systems have smoother, 
Gaussian-like velocity fields (\ie low S).  Both the Virgo and Coma samples 
show the trend of increasing S with luminosity and a similar ``transition 
luminosity''.  There is a hint that the lower luminosity systems in Coma 
may show more structure than do comparable systems in Virgo but the 
inherent uncertainty in our index (S) prevents a more definitive conclusion.

\section{Discussion and Conclusions}

The kinematic structure we find within the cores of luminous 
elliptical galaxies is inconsistent with the smooth, Gaussian profiles 
expected from classical models for elliptical galaxies (\eg Binney 
\& Tremaine 1987).  One possible explanation is that it is produced by 
``stellar streaming''.  However, there may be a problem with the time 
scale.  If we estimate the stellar orbital periods within the cores of 
luminous ellipticals as their scale ($\sim$ 10 kpc) 
divided by the velocity dispersion ($\sim 300$ km/sec), we obtain a 
typical period of $\sim 10^7$ years.  Since phase mixing is expected 
to dilute coherence over only a few orbital periods we estimate the 
lifetime of such stellar streams to be only a few $\times 10^7$ years.  
For comparison the typical crossing time of the galaxies within these 
clusters is much longer, $\sim 10^9$ years.  As a result, it seems unlikely 
that galaxy-galaxy encounters are sufficiently frequent to maintain the 
kinematic structure which we find.

While there may be alternative scenarios which could produce kinematic 
structure, we favor a merger scenario.  Faber \etal (1997) recently 
suggested that the existence of multiple Black Holes (BH) could explain 
the distinction between the photometric profiles of elliptical galaxies.  
The distended cores for the most 
luminous systems can be maintained if the BHs presumed to be present 
in the nuclei of lower luminosity galaxies (\eg van der Marel 1999) 
survive the merger process and then heat the stellar distribution through 
gravitational encounters (\eg Quinlan \& Hernquist 1997).  Faber \etal 
(1997) found that the ``break radius'' which separates the inner 
core profile and the outer $R^{1/4}$ profile correlated strongly with 
galaxy luminosity; the most luminous systems had the largest cores.  
Interestingly, the luminosity at which we find the kinematic structure 
to increase is essentially the same as that found by Faber \etal (1997) 
for the transition between ellipticals with cores and those with cusps.  
We interpret this as further evidence that these two phenomena are related.  
In the context of their model, the kinematic structure which we find 
might be explained as the ``stellar wakes'' produced by these 
BHs as they traverse through the background distribution of stars.  
We speculate that the increased structure we find within the most 
luminous systems could reflect a larger number of merger events 
for these galaxies.  If so, an investigation of this phenomena in field 
environments might be warranted.  Furthermore, the development of the 
NGST could provide the opportunity of investigating this phenomena at 
redshifts of $\sim$ 0.5 and perhaps allow the merger history of luminous 
elliptical galaxies to be quantified.

\section{Summary}

We find evidence for kinematic structure within the cores of luminous 
elliptical galaxies.  Lower luminosity systems show smooth, Gaussian 
velocity fields while the higher luminosity systems show a more irregular 
velocity field.  The structure within luminous systems is most prominent 
within 100 pc of the center with smooth Gaussian velocity fields more 
characteristic of the integrated light within 1 kpc.  We define a 
structure index which is found to be well correlated with the absolute 
magnitude of the galaxy.  We favor an interpretation of this phenomena 
as resulting from coherent stellar streaming and suggest that it could 
be a relic signature of the merger history of luminous elliptical 
galaxies.  There have been recent suggestions that luminous ellipticals 
may harbor several massive Black Holes which heat the stars and produce 
the extended luminosity profiles found in these systems.  In this scenario, 
the kinematic substructure reported here could be interpreted as the 
``stellar wakes'' produced by these Black Holes.  

\section{Acknowledgments}
We wish to thank Sam Barden (NOAO), Dave Sawyer (WIYN), and Kent Honeycutt 
(Indiana) for the development of DensePak.  Robin Tripoli assisted with some 
of the observations described in this paper.

\clearpage

\begin{figure}[ht]
\plotfiddle{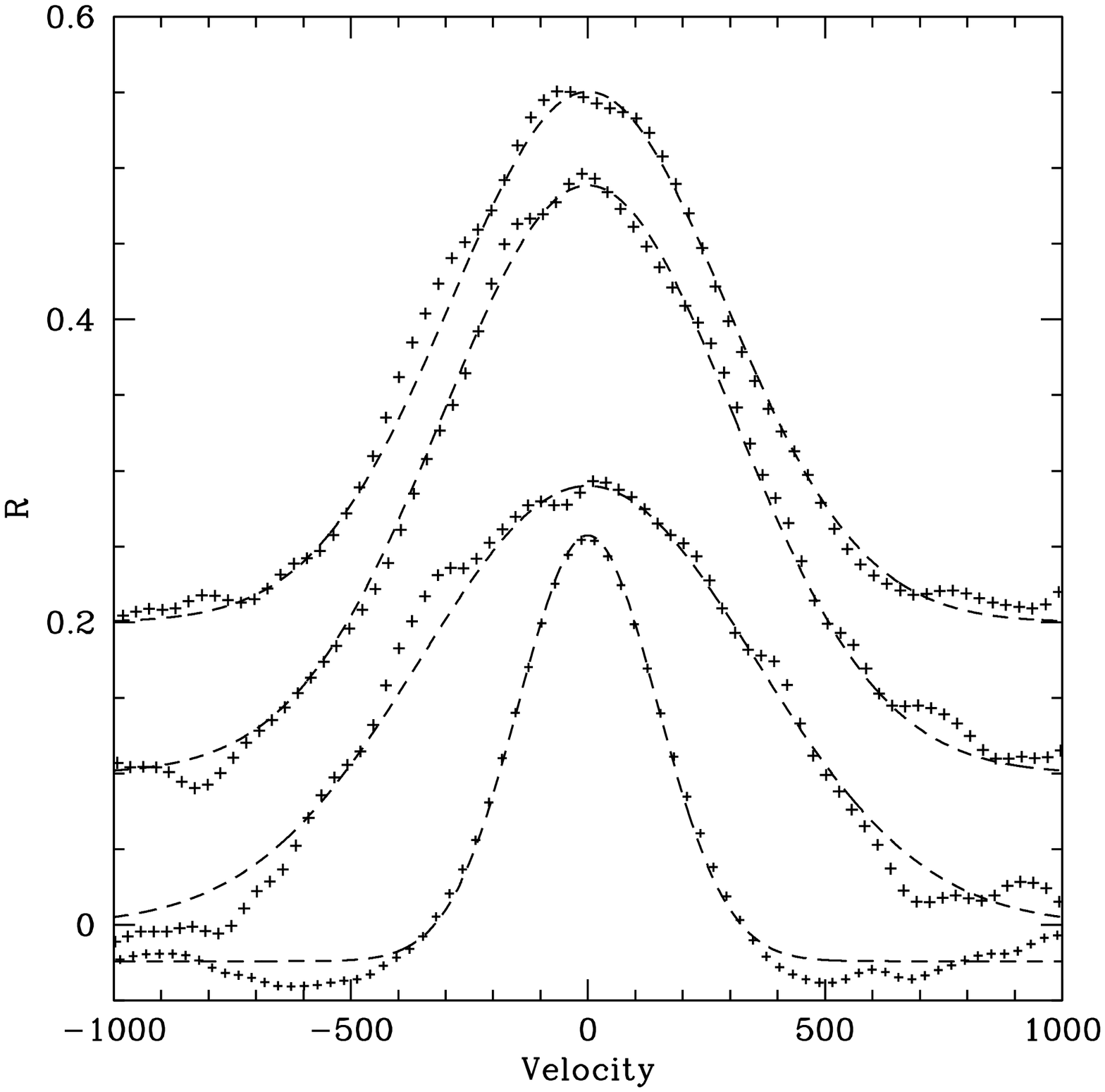}{6.25in}{0}{95}{90}{-310}{-145}
\caption{
Cross-correlation profiles of selected elliptical galaxies in the 
Virgo and Coma clusters.  From top to bottom are shown NGCs 4874 (Coma), 
4472 (Virgo), 4486 (Virgo), and TT 41 (Coma).  The top two have been 
offset by 0.1 in the vertical direction while the bottom plot has been 
offset by -0.05 and scaled by 0.5.  The vertical scale corresponds to 
NGC 4486 (M87) in Virgo.  The crosses are the data and the dashed line 
is the best-fit Gaussian model produced by the IRAF task FXCor.  A 
``structure index'' is defined by counting the number of deviations from 
the best-fit Gaussian (see text and Figure 2).
}\label{fig1}
\end{figure}

\begin{figure}[ht]
\plotfiddle{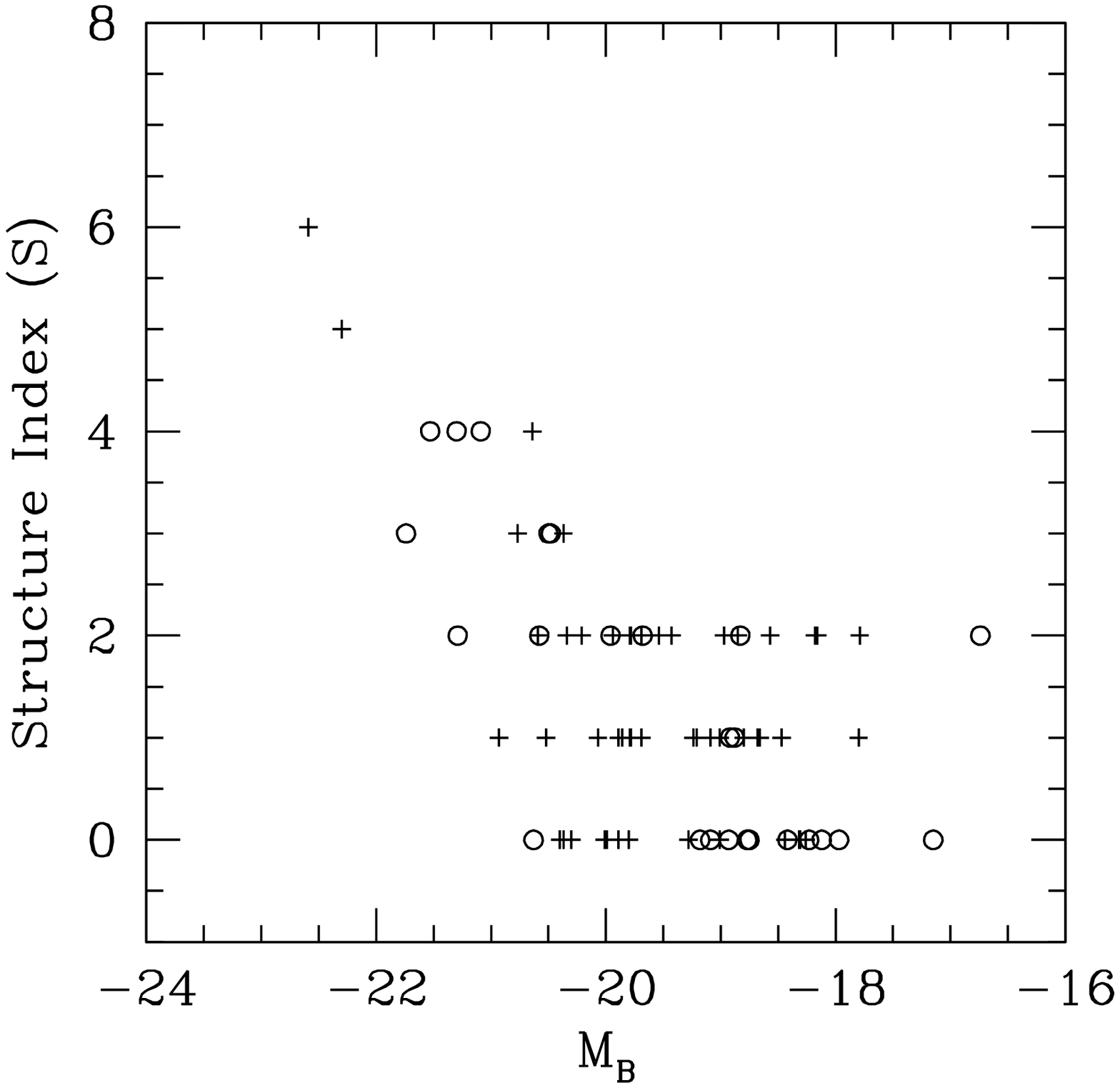}{6.25in}{0}{95}{90}{-340}{-145}
\caption{
The ``structure index'' (see text) as a function of galaxy luminosity.  
Open points are Virgo galaxies assuming $(m-M)_{Virgo} = 31.0$ and crosses 
are Coma cluster galaxies assuming $(m-M)_{Coma} = 34.9$.  Evidently, 
luminous elliptical galaxies within both clusters show an increasing 
amount of kinematic structure with luminosity.
}\label{fig2}
\end{figure}


\clearpage

\begin{references}

\reference{abr}
Barden, S.C., Sawyer, D.G., \& Honeycutt, R.K. 1998, in Optical Astronomical 
Instrumentation, SPIE Vol. 3355, 892

\reference{bin}
Binney, J. \& Tremaine, S. 1987, Galactic Dynamics (Princeton: Princeton 
University Press)

\reference{dev}
de Vaucouleurs, G., de Vaucouleurs, A., Corwin, H.G. Jr., Buta, R.J., 
Paturel, G., \& Fouqu\'e, P. 1991, Third Reference Catalogue of Bright 
Galaxies (New York: Springer)

\reference{djo}
Djorgovski, S., \& Davis, M. 1987, \apj, 313, 59

\reference{dre}
Dressler, A. \etal 1987, \apj, 313, 47

\reference{fab1}
Faber, S.M. \etal 1997, \aj, 114, 1771

\reference{fab2}
Faber, S. M., Burstein, D., Friel, E., \& Gaskell, C. M., 1985, \apjs, 57, 711

\reference{kor}
Kormendy, J., \& Djorgovski, S. 1989, Annual Reviews of Astronomy \& 
Astrophysics, 27, 235

\reference{quin}
Quinlan, G. D., \& Hernquist, L. 1997, New Astronomy, 2, 533

\reference{ton}
Tonry, J.L., \& Davis, M. 1979, \aj, 84, 1511

\reference{tom}
Toomre, A. 1977, in The Evolution of Galaxies and Stellar Populations, 
IAU Symp., No. 58, ed. J.R. Shakeshaft (Dordrecht: Reidel), p. 347.

\reference{van}
van der Marel, R. 1999, \aj, 117, 744

\end{references}
\end{document}